\documentclass[reprint,prl,aps,longbibliography,nofootinbib,floatfix,nobalancelastpage,superscriptaddress]{revtex4-2}
\usepackage{amsmath}
\usepackage{amsfonts}
\usepackage{amssymb}
\usepackage{mathtools}
\usepackage{graphicx}
\usepackage[dvipsnames]{xcolor}
\usepackage{yfonts}
\usepackage{bm}
\usepackage[normalem]{ulem}
\usepackage{dsfont}
\usepackage{braket}
\usepackage{mathrsfs}
\usepackage{upgreek}
\usepackage{tikz}
\usepackage{wasysym}
\usetikzlibrary{decorations.pathmorphing}

\definecolor{myBlue}{RGB}{31,119,180}
\definecolor{myOrange}{RGB}{255,127,14}
\definecolor{myGreen}{RGB}{44,160,44}
\definecolor{myRed}{RGB}{214,39,40}
\definecolor{myPurple}{RGB}{148,103,189}

\usepackage[colorlinks=True,citecolor=Blue,linkcolor=myRed,urlcolor=Blue]{hyperref}
\usepackage{hypcap}

\newcommand{\swap}{\mathsf{SWAP}}

\newcommand\eq[1]{\begin{align}#1\end{align}}

\newcommand\cbox[1]{\vcenter{\hbox{#1}}}

\newcommand{\cc}{{\cal C}}

\begin{document}

\title{On the emergence of quantum many-body chaos for tunably-broken integrability}

\author{Sounak Biswas}
\email{sounak.biswas@icts.res.in}
\affiliation{International Centre for Theoretical Sciences, Tata Institute of Fundamental Research, Bengaluru 560089, India}

\author{Sthitadhi Roy}
\email{sthitadhi.roy@icts.res.in}
\affiliation{International Centre for Theoretical Sciences, Tata Institute of Fundamental Research, Bengaluru 560089, India}

\author{Roderich Moessner}
\email{moessner@pks.mpg.de}
\affiliation{Max-Planck-Institut f\"{u}r Physik komplexer Systeme, N\"{o}thnitzer Stra{\ss}e 38, 01187 Dresden, Germany}

\begin{abstract}
We develop a quantitative theory for the emergence of quantum many-body chaos as integrability is broken via a tunable parameter. 
In a circuit model of free fermions, `doped' with a tunable density of integrability-breaking gates,  we uncover the microscopic mechanisms underpinning the crossover from early-time integrable behaviour to late-time chaos through the lens of the out-of-time-ordered correlators (OTOCs). The integrability-breaking gates act as local, in spacetime, hotspots which locally amplify the OTOCs such that an 
accumulation of them eventually leads to fully-developed chaos. 
We identify the explicit characteristic time and length scales governing this crossover, as well as the dependence of the chaotic OTOC characteristics -- such as the butterfly velocity and front broadening -- on the integrability-breaking parameter. 
\end{abstract}

\maketitle

\paragraph{Introduction:}The emergence of chaos in quantum many-body dynamics is a cornerstone of modern statistical and condensed matter physics. 
It has acquired a newfound significance in the wake of the new generation of quantum simulation platforms which offer unprecedented quantum coherence times and provide access to hitherto unexplored dynamical regimes~\cite{preskill2018quantum,boixo2018characterising,smith2019simulating,arute2019quantum,ippoliti2021many,hoke2023measurement,fauseweh2024quantum}. 
This has naturally spurred the development of new theoretical models to study chaotic dynamics, yielding several exactly soluble settings. 
While these theoretical settings have proven to be immensely insightful, their solubility often comes at the cost of fine-tuning that is difficult to realise experimentally. 
Prominent examples include random circuits or spatially-random Hamiltonians with large local Hilbert spaces~\cite{nahum2017quantum,nahum2018operator,rakovszky2018diffusive,keyserlingk2018operator,khemani2018operator,chan2018solution,chan2018spectral,zhou2019emergent,nahum2022real,fisher2023random,chalker2025chaotic,tan2025operator,tan2026operator} which are manifestly removed from small local Hilbert-space dimensions realised on modern platforms, and dual-unitary circuits~\cite{bertini2018exactSFF,bertini2019entanglementspreading,bertini2019exactcorrelations,claeys2020maximumvelocity,bertini2026exactly-rmp} where the solubility is fragile to generic dual-unitarity breaking perturbations.
Furthermore, the fine-tuned nature of these models leads to them being close to maximally chaotic from the start.

Ideally, a `generic' yet controlled theoretical setting should have as ingredients finite local Hilbert-space dimensions, dynamics driven by local unitary operators capable of forming a universal gate-set, and, a tunable proximity to integrability.
These ingredients are naturally expected to lead to crossover physics from early-time emergent integrable behaviour to fully developed chaos at late times~\cite{bilitewski2021classical,ruidas2026how}. This dynamical regime is not only particularly interesting, not least on account of how common emergent integrability is in physical systems, but also ubiquitously accessible on modern experimental platforms. 
At the same time, this regime also raises fundamental questions about what, possibly generic, mechanisms underpin this crossover and what are the associated time and length scales, and their dependence upon  the integrability-breaking parameter.

\begin{figure}
\includegraphics[width=\linewidth]{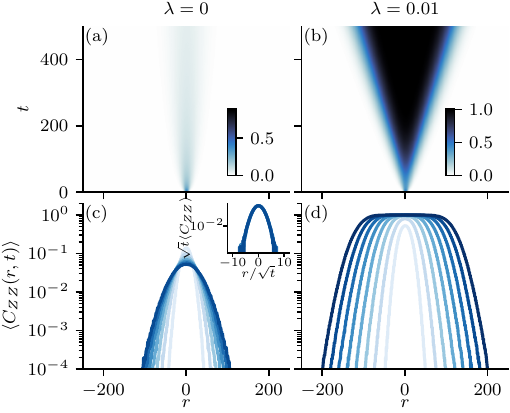}
\caption{The OTOC between the Pauli operators $Z_0$ and $Z_r$, defined in Eq.~\eqref{eq:ZZ-OTOC}. (a) and (b) show the OTOCs as space-time heatmaps for $\lambda=0$ and 0.01 respectively whereas (c) and (d) show time slices for the corresponding OTOCs at $t=0,\cdots,512$ in steps of 64 (lighter to darker colours). At $\lambda=0$, the model is free-fermion integrable leading to a diffusive decay and spreading of the OTOC (as confirmed by the scaling collapse in the inset to (c)), whereas at finite $\lambda$, the model is chaotic, indicated by the ballistic spreading of the OTOC. }
\label{fig:CZZ-lightcone}
\end{figure}

Here we present such a model, for which we study the emergence of chaos through the lens of out-of-time-ordered correlators (OTOCs)~\cite{roberts2015diagnosing,maldacena2016bound,aleiner2016microscopic,bohrdt2017scrambling,luitz2017information}. This is a one-dimensional chain of qubits, or equivalently Majorana fermions. The dynamics is effected by a family of quantum circuits with nearest-neighbour gates with a parameter $\lambda$ which tunes the circuits from being trivially integrable at $\lambda=0$ to chaotic at $\lambda>0$. 
Averaging over the family of circuits allows us to obtain statistically exact results for the OTOCs in arbitrarily large systems. 
Because this framework is robust across the entire range of $\lambda$, we can obtain a quantitative theory of the crossover from integrability to fully developed chaos. 
Specifically, for $\lambda=0$, the circuits comprise only matchgates --- unitary gates which can be expressed via operators quadratic in fermions --- rendering the limit trivially integrable~\cite{valiant2002quantum,terhal2002classical,jozsa2008matchgates}. 
This integrability is broken by doping the circuit with $\swap$ gates, of density $\lambda$, that generate quartic-in-fermions interactions, a setting physically natural also for strongly-correlated condensed matter systems.
Given that matchgates and $\swap$ gates together form a universal gate set~\cite{jozsa2008matchgates}, and our model is defined on qubits with local Hilbert-space dimension of 2, it has all the above mentioned desired ingredients for a `generic' locally interacting quantum many-body system.
Our analysis is rooted in an exact classical Markov process for the averaged OTOCs. 
A continuum description of this yields an effective noisy travelling wave equation for the OTOC of the Fisher-KPP type~\cite{fisher1937wave,kpp1991english}, where $\lambda$ controls the strength of the non-linearities and the noise.

\paragraph{Main results:}At $\lambda=0$, the OTOC simply mimics the dynamics of free particles in a noisy environment, thus exhibiting a Gaussian profile which spreads {\it diffusively},
\eq{
    {\rm OTOC}(r,t;\lambda=0) \approx \frac{1}{\sqrt{2\pi Dt}}\exp\left(-\frac{r^2}{2Dt}\right)\,,
    \label{eq:OTOC-gaussian}
}
At finite $\lambda$, the integrability-breaking gates introduce non-linearities, seeding hotspots of chaotic behaviour by locally amplifying the OTOC. 
At late times, these effects 
accumulate
and the average OTOC spreads {\it ballistically} with a butterfly velocity $v_B$, accompanied by a diffusive broadening of the front.
In this regime it is described by a scaling form
\eq{
{\rm OTOC}(r,t)\!= \!F\!\left(\frac{|r|-v_B(\lambda)t}{\sqrt{t}}\right)\!;
    F(x)\!=\!\begin{cases}
                1;&\!\!\!\!x\!\ll\!1\\
                0;&\!\!\!\!x\!\gg\!1
             \end{cases}\,,
    \label{eq:OTOC-scaling}
}
where $v_B(\lambda)$ scales as $\sim \sqrt{\lambda}$ with logarithmic corrections.
The form above also implies that the diffusion constant associated with the broadening has a negligible dependence on $\lambda$.
The qualitatively different OTOCs between the integrable and chaotic cases are summarised in Fig.~\ref{fig:CZZ-lightcone}.

Crucially, our analysis reveals explicit crossover scales en route to chaos.
We find that that there exists a characteristic timescale, ${t_\ast(\lambda)\sim \lambda^{-1}}$, and a length scale, ${r_\ast(\lambda)\sim \lambda^{-1/2}}$, governing the crossover.
This allows us to identify $r$- and $\lambda$-dependent crossover timescales as 
\eq{
    t_{\rm crossover}(r,\lambda)\sim 
            \begin{cases}
                t_\ast(\lambda)\,;~~~& r \ll r_\ast(\lambda)\\
                \frac{r}{r_\ast(\lambda)}t_\ast(\lambda)\,;~~~ & r\gg r_\ast(\lambda)
            \end{cases}\,.
}
The following physical picture therefore emerges.
For $r\ll r_\ast(\lambda)$, the diffusively spreading OTOC of the integrable limit leads to a local temporal maximum before the `chaotic' OTOC front, seeded by the $\swap$\ gates, arrives at the crossover timescale set by $t_\ast(\lambda)$.
By contrast, at $r\gg r_\ast(\lambda)$, the OTOC front that arrives is already chaotic as the timescale of arrival is much larger than $t_\ast(\lambda)$. The crossover timescale is therefore just the arrival time, $\sim r/v_{B}(\lambda)$. Since $v_B\sim\sqrt{\lambda}$, the crossover timescale scales effectively as $t_\ast(\lambda)\times r/r_\ast(\lambda)$.

In the remainder of the paper, we first concretely define the model of $\swap$-doped matchgate circuits and the OTOCs we compute.
We then discuss the classical Markov process for the averaged OTOC and present numerical results obtained from it, for both the crossover and chaotic regimes.
This is followed by analytical arguments for the results based on a continuum, noisy F-KPP equation obtained from the classical Markov process. 

\paragraph{Definition of OTOCs:} Denoting the Pauli matrices for the qubit (at site $x$) by $\{X_x,Y_x,Z_x\}$, we define the OTOC as 
\eq{
    C_{ZZ}(r,t) = \frac{1}{2}{\rm tr}\left([Z_0(t),Z_r][Z_0(t),Z_r]^\dagger\right)\,,
    \label{eq:ZZ-OTOC}
}
We will also find it useful to consider the language of Majorana fermions. 
Denoting the two Majoranas on a qubit site as $\gamma_x^{A} = \left(\otimes_{x'=1}^{x-1}Z_{x'}\right)X_x$ and $\gamma_x^{B} = \left(\otimes_{x'=1}^{x-1}Z_{x'}\right)Y_x$, we define the Majorana OTOC as 
\eq{
    C^{\mu\nu}(r,t) = \frac{1}{2}{\rm tr}\left[\{\gamma_0^\mu(t),\gamma_{r}^\nu\}\{\gamma_0^\mu(t),\gamma_r^\nu\}^\dagger\right]\,,
        \label{eq:majo-OTOC}
}
where $\{\cdot,\cdot\}$ denotes the anticommutator.
In particular, it will be convenient to write the time-evolving operators in the basis of $4^L$ Majorana strings ${\cal S}$ as 
\eq{
    Z_0(t) = \sum_{{\cal S}}c_{{\cal S}}(t){\cal S}\,;~\gamma_0^\mu(t) = \sum_{{\cal S}}d_{{\cal S}}(t){\cal S}\,,
}
where a Majorana string can be represented as ${\cal S} = \otimes_{x=1}^L (\gamma_x^A)^{\eta_{x,{\cal S}}^A}(\gamma_x^B)^{\eta_{x,{\cal S}}^B}$ with 
$\eta_{x,{\cal S}}^{\nu}=0,1$ indicating if the operator $\gamma_x^{\nu}$ is absent or present in the string ${\cal S}$. 
Using this notation, the OTOCs in Eq.~\eqref{eq:ZZ-OTOC} and Eq.~\eqref{eq:majo-OTOC} can be written as 
\eq{
	\begin{split}
		C_{ZZ}(r,t) &= 2\sum_{{\cal S}}|c_{{\cal S}}(t)|^2 (\eta^A_{r,{\cal S}}+\eta^B_{r,{\cal S}}-2\eta^A_{r,{\cal S}}\eta^B_{r,{\cal S}})\,,\\
		C^{\mu\nu}(r,t) &= 2\sum_{{\cal S}}|d_{{\cal S}}(t)|^2 \eta_{r,{\cal S}}^\nu\,.
	\end{split}
	\label{eq:OTOCs-occ}
}
Physically, this implies that the $ZZ$-OTOC at $(r,t)$ is given by the total weight of all those Majorana strings in the time-evolving operator which contain 
exactly one Majorana 
at site $r$ and time $t$.
Similarly, the Majorana OTOC in Eq.~\eqref{eq:majo-OTOC} is given by the total weight of those strings which contain $\gamma_r^\nu$.

\paragraph{Model:}
We consider a brickwork circuit where the time-evolution operator from time $t$ to $t+1$ is given by a layer of gates acting on the odd bonds followed by a layer of gates acting on the even bonds,
\eq{
    {\cal U}_t = \bigotimes_{x=1}^{L/2-1}U_{2x,2x+1}^{(t)}\bigotimes_{x=1}^{L/2}U_{2x-1,2x}^{(t)}\,.
}
The gate acting on sites $x$ and $x+1$ at time $t$ is
\eq{
    U^{(t)}_{x,x+1}\!=\!\begin{cases}
                        M_{x,x+1}^{(t)}\,;&p=1-\lambda\\
                        S_{x,x+1}^{(t)}\,;&p=\lambda\\
                        \end{cases}\,,
    \label{eq:brickwork}
}
where ${S_{x,x+1}^{(t)}=M_{x,x+1}^{(t)}\swap_{x,x+1}\tilde{M}_{x,x+1}^{(t)}}$ with $M_{x,x+1}^{(t)}$ a randomly chosen matchgate\footnote{The $\swap_{x,y}$ gate, given by $(\mathbb{I}_x\mathbb{I}_y+X_xX_y+Y_xY_y+Z_xZ_y)/2$, swaps the states of the qubits.}.
An instance of a matchgate is given by two independent $2\times2$ unitary matrices, $u$ and $v$, each of which acts within a single parity sector in the $Z$ basis, subject to the constraint that both of them have the same determinant. 
In the following, we  sample matchgates by sampling $u$ and $v$ uniformly from $SU(2)$.
Given that a matchgate acting on sites $x$ and $x+1$ can be expressed as a sum of operators quadratic in the fermions, evolving any Majorana string supported on the two sites by it leads to a superposition of strings each operator in which has exactly the same number of Majorana operators as the input string. For instance,
\eq{
	M_{x,x+1}\gamma_x^\mu M_{x,x+1}^\dagger = \sum_{\nu=A,B}[a_{x}^{\nu}\gamma_x^\nu + a_{x+1}^\nu\gamma_{x+1}^\nu]\,.
	\label{eq:mg-single-majorana}
}
On the other hand, the $\swap$ gates can expand or contract the Majorana string into a linear combination of strings where each term can be a product of more Majorana operators than the input. For instance, the SWAP acting on a single Majorana operator leads to a sum of 3-Majorana operators, 
\eq{
	S_{x,x+1}\gamma_x^\mu S_{x,x+1}^\dagger = \sum_{\nu=A,B}[&b_{x,x+1}^{AB;\nu}\gamma_x^A\gamma_x^B\gamma_{x+1}^\nu + \nonumber\\&b_{x,x+1}^{\nu;AB}\gamma_{x}^\nu\gamma_{x+1}^A\gamma_{x+1}^B]\,,
	\label{eq:sw-single-majorana}
}
which is a direct manifestation of the fact that the $\swap$ gate induces quartic-in-fermions interaction terms. 

\paragraph{Classical Markov process:}
The dynamics of the weight of the strings in Eq.~\eqref{eq:OTOCs-occ}, averaged over the ensemble of random matchgates, can be mapped exactly onto a classical Markov process, which we discuss now.
The key point is that the averaging over the matchgates twirls the operators strings such that the average weight of any of the possible resultant strings is equal. 
In the example in Eq.~\eqref{eq:mg-single-majorana}, this implies $\braket{|a_{x(+1)}^\nu|^2}_{M} = 1/4$ where $\braket{\cdot}_M$ denotes the average over the matchgates.
Similarly, averaging over the matchgates in $S$ in Eq.~\eqref{eq:sw-single-majorana} leads to an equipartition of probabilities over the possible 3-Majorana strings; such that $\braket{|b_{x,x+1}^{AB;\nu}|^2}_M=1/4$.
This leads to the to an effective classical Markov process for the average occupancies of the Majoranas and therefore the average OTOCs.
It is important to note that the rules of the classical Markov process are exact.

To describe the processes concretely, it is useful to introduce a notation $\Gamma_{\cal S}\equiv\{\eta_{x,\cal S}^\mu\}$, where $\Gamma_{\cal S}$ denotes a configuration of hard-core particles on a chain with $2L$ sites labelled as $\{x^\mu\}$ with $\mu=A,B$ and $x\in (-L/2,L/2]$, and $\eta_{x,\cal S}^\mu=0,1$ denotes the sites to be empty or occupied.
The OTOCs averaged over the matchgates, in this picture, are given by equations of the form \eqref{eq:OTOCs-occ} but with $|c_{\cal S}(t)|^2$ and $|d_{\cal S}(t)|^2$ replaced by the probability $P_{\Gamma_{\cal S}}(t)$ of the configuration $\Gamma_{\cal S}(t)$ in the Markov process.
This effectively means that $\braket{C_{ZZ}(r,t)}$ is given by the probability of having exactly one of $r^A$ and $r^B$ occupied and similarly, $\braket{ C^{\mu\nu}(r,t)}$ is given by the probability of $r^\nu$ being occupied, at time $t$. 
The initial condition for the process is simply having both the sites at $r=0$ occupied for $C_{ZZ}$ and having just $0^\mu$ occupied for $C^{\mu\nu}$.

We next describe the rules of the Markov process in words and present their details in the End Matter (EM).
Consider sites $x$ and $x+1$, a configuration where any $n$ of the four Majorana slots are occupied. 
If the gate encountered by the pair of sites is a matchgate, then resulting configuration is one of the $\binom{4}{n}$ configurations, each with a probability $\binom{4}{n}^{-1}$.
This implies that the Markov processes associated to the matchgates only effect a random walk of the particles subject to the hard-core constraint and their number conservation. 
For the OTOCs, since the initial condition consists of only one or two particles, it is straightforward to conclude that the probability of a finding the particle at $(r,t)$ follows a Gaussian distribution with a standard deviation which scales as $\sqrt{t}$.
This leads to the diffusive spreading of the OTOC mentioned in Eq.~\eqref{eq:OTOC-gaussian} with $D=2$,
and shown in Fig.~\ref{fig:CZZ-lightcone}(a),(c).
The Gaussian profile of the OTOCs, and more importantly, the absence of a butterfly velocity is a clear signature of the absence of chaos in this trivially integrable limit.

At finite $\lambda$, the presence of $\swap$ gates changes the situation qualitatively.
If the pair of sites encounters a $\swap$ gate (sandwiched between matchgates), then the number of occupied slots in the resulting configuration depends on $n$.
For $n=1(3)$, the output configuration has three (one) occupied slots and the probability of any one of such four configurations is $1/4$.
On the other hand, if $n=0,2,4$, then the output configuration continues have the same number of occupied slots; the probability is equipartitioned between the six possible configurations for $n=2$ whereas for $n=0,4$ the configurations are obviously unique.
The key point in the above rules is that the $\swap$ gates can expand a single Majorana into a product of three Majoranas.
In terms of operator growth, it means that a given Majorana operator appears in three different Majorana strings which in turn locally amplifies the OTOC and the $\swap$ gates act as local hotspots for the OTOC.
The matchgates move these particles through the system until the next $\swap$ gate again causes a local growth in the number of particles and concomitantly, in the OTOC.
A sequence of such events, spread over the system, eventually leads to a global amplification of the OTOC and the crossover to the chaotic behaviour -- this is the central mechanism that we identify in this work. 
Averaging over the locations of the $\swap$ gates leads to a ballistic lightcone of the OTOCs with it saturating inside the bulk.

\begin{figure}[!t]
\includegraphics[width=\linewidth]{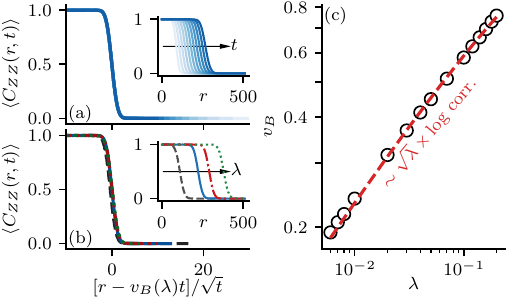}
\caption{Evidence for the scaling form of the OTOC in Eq.~\eqref{eq:OTOC-scaling}. (a) $\braket{C_{ZZ}(r,t)}$ for $\lambda=0.01$ at different $t$ (200 to 600 in steps of 40) collapses onto a common curve when plotted as a function of $(r-v_Bt)/\sqrt{t}$; the inset shows the data just a function of $r$. (b) The data at a fixed $t=600$ but for different values of $\lambda = 0.01, 0.05,0.1,0.2$ again collapses onto a common curve indicating that the diffusion constant associated to the broadening of the front is independent of $\lambda$. (c) $v_B$ extracted from the scaling collapses shows a behaviour with $\lambda$ consistent with $\sqrt{\lambda}$ modulo logarithmic corrections.}
\label{fig:vb-collapse}
\end{figure}

In Fig.~\ref{fig:vb-collapse}, we present numerical evidence for the result in Eq.~\eqref{eq:OTOC-scaling}; plotting the OTOC as a function of $[r-v_B(\lambda)t]/\sqrt{t}$ collapses the data for different $t$ [panel (a)] as well as different $\lambda$ [panel (b)] onto a common curve.
The $v_B(\lambda)$ extracted from the data collapse is shown as a function of $\lambda$ in panel (c). 
While the data points appear to fall on a straight line, indicating a power-law, the data are in fact consistent with $v_{B}\sim \sqrt{\lambda}$ with logarithmic corrections, which is predicted from our theory as we will discuss shortly.

\begin{figure}[!t]
\includegraphics[width=\linewidth]{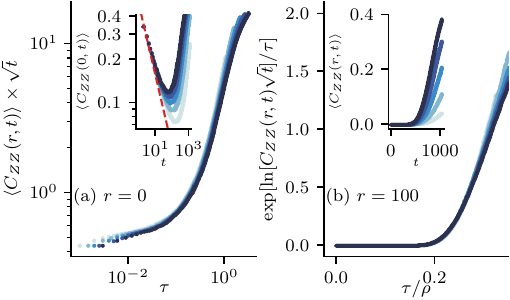}
\caption{Evidence for the crossover scaling in Eq.~\eqref{eq:OTOC-crossover-scaling} with ${t_\ast(\lambda)\sim\lambda^{-1}}$ and ${r_\ast(\lambda)\sim \lambda^{-1/2}}$. (a) The OTOC at $r=0$ rescaled by $\sqrt{t}$, for different $\lambda$ collapses onto a common curve as a function of $\tau=t\lambda$; the inset shows that the OTOC follows the $1/\sqrt{t}$ decay (red dashed line) of the integrable limit until $t_\ast(\lambda)$. (b) Similar analysis but for $r=100$ again shows that the OTOC rescaled by $\sqrt{t}$ is a function of $\rho(\lambda)=r\sqrt{\lambda}$ and $\tau$. Lighter to darker colours indicate $\lambda = 10^{-3}$ to $3\times10^{-3}$ in steps of $5\times 10^{-4}$ and data are for $L=1024$.  }
\label{fig:crossover}
\end{figure}

We now turn towards the crossover to chaotic behaviour 
in Eq.~\eqref{eq:OTOC-scaling} from the integrable behaviour in Eq.~\eqref{eq:OTOC-gaussian}.
The results in Fig.~\ref{fig:crossover} suggest the existence of characteristic crossover timescales and length scales,
$t_\ast(\lambda)\sim \lambda^{-1}\,;~r_\ast(\lambda)\sim \lambda^{-1/2}$,
such that the OTOC in terms of rescaled variables, $\rho(\lambda) = r/r_\ast(\lambda)$ and $\tau(\lambda)=t/t_\ast(\lambda)$, satisfies a scaling form
\eq{
    \braket{C_{ZZ}(r,t)} = t^{-1/2}\exp[\tau g(\rho/\tau)]\,.
    \label{eq:OTOC-crossover-scaling}
}
Note that the OTOC is a function of two variables, $r$ and $t$ and hence the scaling form in Eq.~\ref{eq:OTOC-crossover-scaling} is a function of tw variables $\rho$ and $\tau$.
In Fig.~\ref{fig:crossover} we show two representative spatial cuts, one at $r=\rho=0$ (a) and the other at $r=100$ and the corresponding $\rho$ values (b).
For completeness, in the EM, we also show scaling collapses to the form in Eq.~\ref{eq:OTOC-crossover-scaling} along representative temporal cuts.
These results provide numerical evidence for the scaling of $t_\ast(\lambda)$ and $r_\ast(\lambda)$ with $\lambda$.

\paragraph{Noisy F-KPP description:} To get analytical insights into the said crossover scales as well as the chaotic behaviour of the OTOC, we turn to a continuum description of the Markov process. 
For convenience, we consider the  Majorana OTOC.
The coarse-grained OTOC, ${\cal C}(r,t)$ is described by a noisy F-KPP equation~\cite{fisher1937wave,kpp1991english},
\eq{
    \partial_t {\cal C} = D\partial_r^2{\cal C} + \lambda H[{\cal C}] + \sqrt{\lambda K[{\cal C}]}\eta(r,t)\,,
    \label{eq:FKPP}
}
where $H[{\cal C}] = 2{\cal C}(1-{\cal C})(1-2{\cal C})$, $K[{\cal C}] = {\cal C}(1-{\cal C})(1-2{\cal C}+2{\cal C}^2)$, and $\eta(r,t)$ is white noise.
Equations of the F-KPP form have been discussed in the context of OTOCs in models of weakly interacting electrons and random circuits with large local Hilbert spaces~\cite{aleiner2016microscopic,nahum2018operator,zhou2023hydrodynamic,narde2026entanglement,swann2026continuum}, and lead to travelling-wave solutions for the OTOC with a butterfly velocity and diffusively broadening front.
We derive it in our case from the microscopic rules of the Markov process.
We relegate the details of the derivation to the EM and only discuss here the physical origins of the different terms.

The first term is the diffusion of the particles effected by the matchgates.
The second term encodes the non-linearities due to the $\swap$ gates and hence appears with a rate $\lambda$. 
At a pair of sites, the probability of there being just a one and three particles is $p_1\sim \cc(1-\cc)^3$ and $p_3 \sim \cc^3(1-\cc)$. 
Branching and annihilation due to the $\swap$ gates in the two cases leads to amplification and decay respectively of the local particle number by 2.
The nonlinear term is therefore $H[\cc]\sim 2(p_1-p_3)$ which is exactly the form above. 
While $H[\cc]$ the captures mean-field operator growth, it implicitly assumes a continuous fluid where $\cc$ can take arbitrarily small values. This is at odds with a strictly finite local Hilbert-space dimension which imposes a natural but random cutoff on the front's position leading to the demographic noise, $\sqrt{\lambda K[{\cal C}]}$~\cite{brunet1997shift}, which leads to a diffusive broadening of the front. 

In the bulk of the lightcone, the effect of this noise can be neglected.
Furthermore, in the crossover regime, ${\cal C}\ll 1$ which justifies a linear approximation, $H[{\cal C}]\sim 2{\cal C}$ in Eq.~\eqref{eq:FKPP} such that the OTOC has a solution
\eq{
    {\cal C}(r,t) \approx \exp[2\lambda t(1-r^2/8\lambda t^2)]/\sqrt{4\pi t}\,.
    \label{eq:linearised-soln}
}
This naturally yields a characteristic timescale
\eq{
    t_\ast(\lambda)\sim (1 /4\lambda)\ln(1/{4\lambda}) + \cdots\,,
}
at which the OTOC  at $r=0$ feels the presence of the $\swap$ gates and crosses over from the integrable, diffusive decay to a chaotic growth, and thus explains the result in Fig.~\ref{fig:crossover}(a).

At any finite $r$, the OTOC in the integrable limit exhibits a local maximum in time before decaying asymptotically as $t^{-1/2}$. 
The length scale $r_\ast(\lambda) \sim 1/\sqrt{\lambda}$ demarcates two distinct spatial regimes for the onset of chaos. 
For $r \ll r_\ast(\lambda)$, chaotic behaviour emerges only after the OTOC has traversed this integrable local maximum. 
Conversely, for $r \gg r_\ast(\lambda)$, the OTOC departs from the integrable baseline before any such maximum is reached. 
At the leading-order mean-field level, the crossover scale $r_\ast(\lambda)$ can be estimated by determining the distance $r$ up to which Eq.~\eqref{eq:linearised-soln} continues to support a local temporal maximum.
This results in  $r_\ast(\lambda)\sim 1/\sqrt{\lambda}$, consistent with the numerical results in Fig.~\ref{fig:crossover}(b).

In the regime of fully-developed chaos, the noisy F-KPP equation predicts the butterfly velocity to be of the form
$v_B\sim \sqrt{\lambda}[1-a/(|\ln \lambda|+b)^2]$~\cite{brunet1997shift,brunet2006phenomenological} where $a,b$ are constants.
This is the functional form of the red dashed line in Fig.~\ref{fig:vb-collapse}(c) which is  in good agreement with the numerical data, although it is hard to rule out other possible functional forms given how featureless the curve is overall.
In addition, by a change of coordinates, $r\to\rho(\lambda)$ and $t\to\tau(\lambda)$, in Eq.~\eqref{eq:FKPP}, it is straightforward to show that the diffusion constant associated to the broadening of the OTOC front is independent of $\lambda$ (modulo logarithmic corrections, see EM), consistent with the numerical results in Fig.~\ref{fig:vb-collapse}.
This concludes our demonstration of how the noisy F-KPP framework captures analytically all the aspects of the OTOC including the crossover behaviour as well as the saturation at chaos.

\paragraph{Outlook:} In this work, we have quantitatively studied the emergence of chaos for tunably-broken integrability through the lens of OTOCs. 
A natural generalisation would be to extend the framework for higher-order $k$-OTOCs, which probe the emergence of free independence and quantum designs~\cite{roberts2017chaos,fava2025designs,trigueros2026unitary}.
In particular, it would be interesting to investigate if there exists a hierarchy of crossover time and length scales with $k$, alongside their dependence on the integrability-breaking parameter.

\begin{acknowledgements}
We thank S. Bhattacharjee, J. T. Chalker and A. Kundu for useful discussions. 
S.B. is supported by the Swarna Jayanti fellowship grant of SERB-DST (India) Grant No. SB/SJF/2021-22/12.
S.B. and S.R. acknowledge the support of the Department of Atomic Energy, Government of India, under project nos. RTI4019 and RTI4013 as well as a Max Planck Partner Group grant between ICTS-TIFR, Bengaluru and MPIPKS, Dresden.
S.R. acknowledges support from SERB-DST, Government of India, under Grant No. SRG/2023/000858 and ANRF (India) under Grant No. ANRF/ARG/2025/004045/PS. This work was supported in part by the Deutsche Forschungsgemeinschaft via the cluster of excellence ctd.qmat (EXC 2147, project-id 390858490)  and  SFB 1143 (Project-ID No. 247310070).
\end{acknowledgements}

\bibliography{refs}

@article{tan2025operator,
  title = {Operator spreading in random unitary circuits with unitary-invariant gate distributions},
  author = {Tan, Zhiyang and Brouwer, Piet W.},
  journal = {Phys. Rev. B},
  volume = {111},
  issue = {18},
  pages = {184301},
  numpages = {20},
  year = {2025},
  month = {May},
  publisher = {American Physical Society},
  doi = {10.1103/PhysRevB.111.184301},
  url = {https://link.aps.org/doi/10.1103/PhysRevB.111.184301}
}

@misc{trigueros2026unitary,
      title={{Unitary Designs from Doped Matchgate Circuits}}, 
      author={Fabian Ballar Trigueros and Zheng-Hang Sun and Xhek Turkeshi and Piotr Sierant and Poetri Sonya Tarabunga},
      year={2026},
      eprint={2606.23800},
      archivePrefix={arXiv},
      primaryClass={quant-ph},
      url={https://arxiv.org/abs/2606.23800}, 
}

@misc{tan2026operator,
      title={Operator spreading in random circuits with orthogonal or symplectic symmetry}, 
      author={Zhiyang Tan and Piet W. Brouwer},
      year={2026},
      eprint={2606.03956},
      archivePrefix={arXiv},
      primaryClass={quant-ph},
      url={https://arxiv.org/abs/2606.03956}, 
}

@article{roberts2017chaos,
   title={Chaos and complexity by design},
   volume={2017},
   pages={121},
   ISSN={1029-8479},
   url={http://dx.doi.org/10.1007/JHEP04(2017)121},
   number={4},
   journal={J. High Energy Phys},
   publisher={Springer Science and Business Media LLC},
   author={Roberts, Daniel A. and Yoshida, Beni},
   year={2017},
   month=Apr }

@article{fava2025designs,
  title = {{Designs via Free Probability}},
  author = {Fava, Michele and Kurchan, Jorge and Pappalardi, Silvia},
  journal = {Phys. Rev. X},
  volume = {15},
  issue = {1},
  pages = {011031},
  numpages = {26},
  year = {2025},
  month = {Feb},
  publisher = {American Physical Society},
  doi = {10.1103/PhysRevX.15.011031},
  url = {https://link.aps.org/doi/10.1103/PhysRevX.15.011031}
}

@misc{chalker2025chaotic,
      title={Chaotic many-body quantum dynamics, spectral correlations, and energy diffusion}, 
      author={J. T. Chalker and Dominik Hahn},
      year={2025},
      eprint={2510.02198},
      archivePrefix={arXiv},
      primaryClass={quant-ph},
      url={https://arxiv.org/abs/2510.02198}, 
}

@misc{ruidas2026how,
      title={How many-body chaos emerges in the presence of quasiparticles}, 
      author={Sibaram Ruidas and Sthitadhi Roy and Subhro Bhattacharjee and Roderich Moessner},
      year={2026},
      eprint={2601.05238},
      archivePrefix={arXiv},
      primaryClass={cond-mat.stat-mech},
      url={https://arxiv.org/abs/2601.05238}, 
}

@article{nahum2022real,
  title = {Real-time correlators in chaotic quantum many-body systems},
  author = {Nahum, Adam and Roy, Sthitadhi and Vijay, Sagar and Zhou, Tianci},
  journal = {Phys. Rev. B},
  volume = {106},
  issue = {22},
  pages = {224310},
  numpages = {26},
  year = {2022},
  month = {Dec},
  publisher = {American Physical Society},
  doi = {10.1103/PhysRevB.106.224310},
  url = {https://link.aps.org/doi/10.1103/PhysRevB.106.224310}
}

@article{terhal2002classical,
  title = {Classical simulation of noninteracting-fermion quantum circuits},
  author = {Terhal, Barbara M. and DiVincenzo, David P.},
  journal = {Phys. Rev. A},
  volume = {65},
  issue = {3},
  pages = {032325},
  numpages = {10},
  year = {2002},
  month = {Mar},
  publisher = {American Physical Society},
  doi = {10.1103/PhysRevA.65.032325},
  url = {https://link.aps.org/doi/10.1103/PhysRevA.65.032325}
}

@article{jozsa2008matchgates,
  title={Matchgates and classical simulation of quantum circuits},
  author={Jozsa, Richard and Miyake, Akimasa},
  journal={Proc. R. Soc. A},
  volume={464},
  number={2100},
  pages={3089--3106},
  year={2008},
  publisher={The Royal Society London},
  doi={10.1098/rspa.2008.0189}
}

@article{valiant2002quantum,
  title={Quantum circuits that can be simulated classically in polynomial time},
  author={Valiant, Leslie G},
  journal={SIAM Journal on Computing},
  volume={31},
  number={4},
  pages={1229--1254},
  year={2002},
  publisher={SIAM},
  doi={10.1137/070682575}
}

@article{brunet2006phenomenological,
  title = {Phenomenological theory giving the full statistics of the position of fluctuating pulled fronts},
  author = {Brunet, E. and Derrida, B. and Mueller, A. H. and Munier, S.},
  journal = {Phys. Rev. E},
  volume = {73},
  issue = {5},
  pages = {056126},
  numpages = {9},
  year = {2006},
  month = {May},
  publisher = {American Physical Society},
  doi = {10.1103/PhysRevE.73.056126},
  url = {https://link.aps.org/doi/10.1103/PhysRevE.73.056126}
}

@article{brunet1997shift,
  title = {Shift in the velocity of a front due to a cutoff},
  author = {Brunet, Eric and Derrida, Bernard},
  journal = {Phys. Rev. E},
  volume = {56},
  issue = {3},
  pages = {2597--2604},
  numpages = {0},
  year = {1997},
  month = {Sep},
  publisher = {American Physical Society},
  doi = {10.1103/PhysRevE.56.2597},
  url = {https://link.aps.org/doi/10.1103/PhysRevE.56.2597}
}

@article{aleiner2016microscopic,
title = {{Microscopic model of quantum butterfly effect: Out-of-time-order correlators and traveling combustion waves}},
journal = {Ann. Phys.},
volume = {375},
pages = {378-406},
year = {2016},
issn = {0003-4916},
doi = {https://doi.org/10.1016/j.aop.2016.09.006},
url = {https://www.sciencedirect.com/science/article/pii/S0003491616301919},
author = {Igor L. Aleiner and Lara Faoro and Lev B. Ioffe},
}

@article{nahum2018operator,
 title = {{Operator Spreading in Random Unitary Circuits}},
  author = {Nahum, Adam and Vijay, Sagar and Haah, Jeongwan},
  journal = {Phys. Rev. X},
  volume = {8},
  issue = {2},
  pages = {021014},
  numpages = {30},
  year = {2018},
  month = {Apr},
  publisher = {American Physical Society},
  doi = {10.1103/PhysRevX.8.021014},
  url = {https://link.aps.org/doi/10.1103/PhysRevX.8.021014}
}

@article{zhou2023hydrodynamic,
  title = {Hydrodynamic theory of scrambling in chaotic long-range interacting systems},
  author = {Zhou, Tianci and Guo, Andrew and Xu, Shenglong and Chen, Xiao and Swingle, Brian},
  journal = {Phys. Rev. B},
  volume = {107},
  issue = {1},
  pages = {014201},
  numpages = {16},
  year = {2023},
  month = {Jan},
  publisher = {American Physical Society},
  doi = {10.1103/PhysRevB.107.014201},
  url = {https://link.aps.org/doi/10.1103/PhysRevB.107.014201}
}

@misc{swann2026continuum,
      title={Continuum mechanics of entanglement in noisy interacting fermion chains}, 
      author={Tobias Swann and Adam Nahum},
      year={2026},
      eprint={2601.21134},
      archivePrefix={arXiv},
      primaryClass={cond-mat.stat-mech},
      url={https://arxiv.org/abs/2601.21134}, 
}

@misc{narde2026entanglement,
      title={Entanglement spreading and emergent locality in Brownian SYK chains}, 
      author={Jatin Narde and Onkar Parrikar and Harshit Rajgadia and Sandip Trivedi},
      year={2026},
      eprint={2508.00060},
      archivePrefix={arXiv},
      primaryClass={hep-th},
      url={https://arxiv.org/abs/2508.00060}, 
}

@incollection{kpp1991english,
  author    = {Kolmogorov, Andrei N. and Petrovsky, Ivan G. and Piskunov, Nikolai S.},
  title     = {Study of the diffusion equation with growth of the amount of matter and its application to a biological problem},
  booktitle = {Selected Works of A. N. Kolmogorov: Volume I: Mathematics and Mechanics},
  editor    = {Tikhomirov, Vladimir M.},
  translator= {Volosov, V. M.},
  pages     = {242--270},
  publisher = {Kluwer Academic Publishers},
  address   = {Dordrecht},
  year      = {1991},
  series    = {Mathematics and Its Applications (Soviet Series)},
  volume    = {25}
}

@article{fisher1937wave,
  title={The wave of advance of advantageous genes},
  author={Fisher, Ronald Aylmer},
  journal={Annals of eugenics},
  volume={7},
  number={4},
  pages={355--369},
  year={1937},
  publisher={Wiley Online Library},
  doi={10.1111/j.1469-1809.1937.tb02153.x}
}

@article{zhou2019emergent,
  title = {Emergent statistical mechanics of entanglement in random unitary circuits},
  author = {Zhou, Tianci and Nahum, Adam},
  journal = {Phys. Rev. B},
  volume = {99},
  issue = {17},
  pages = {174205},
  numpages = {28},
  year = {2019},
  month = {May},
  publisher = {American Physical Society},
  doi = {10.1103/PhysRevB.99.174205},
  url = {https://link.aps.org/doi/10.1103/PhysRevB.99.174205}
}

@article{bertini2018exactSFF,
  title = {Exact Spectral Form Factor in a Minimal Model of Many-Body Quantum Chaos},
  author = {Bertini, Bruno and Kos, Pavel and Prosen, T.},
  journal = {Phys. Rev. Lett.},
  volume = {121},
  issue = {26},
  pages = {264101},
  numpages = {6},
  year = {2018},
  month = {Dec},
  publisher = {American Physical Society},
  doi = {10.1103/PhysRevLett.121.264101},
  url = {https://link.aps.org/doi/10.1103/PhysRevLett.121.264101}
}

@article{bertini2019entanglementspreading,
  title = {Entanglement Spreading in a Minimal Model of Maximal Many-Body Quantum Chaos},
  author = {Bertini, Bruno and Kos, Pavel and Prosen, T.},
  journal = {Phys. Rev. X},
  volume = {9},
  issue = {2},
  pages = {021033},
  numpages = {27},
  year = {2019},
  month = {May},
  publisher = {American Physical Society},
  doi = {10.1103/PhysRevX.9.021033},
  url = {https://link.aps.org/doi/10.1103/PhysRevX.9.021033}
}

@article{bertini2019exactcorrelations,
  title = {Exact Correlation Functions for Dual-Unitary Lattice Models in $1+1$ Dimensions},
  author = {Bertini, Bruno and Kos, Pavel and Prosen, T.},
  journal = {Phys. Rev. Lett.},
  volume = {123},
  issue = {21},
  pages = {210601},
  numpages = {6},
  year = {2019},
  month = {Nov},
  publisher = {American Physical Society},
  doi = {10.1103/PhysRevLett.123.210601},
  url = {https://link.aps.org/doi/10.1103/PhysRevLett.123.210601}
}

@article{bertini2026exactly-rmp,
  title = {Exactly solvable quantum many-body dynamics from space-time duality},
  author = {Bertini, Bruno and Claeys, Pieter W. and Prosen, T.},
  journal = {Rev. Mod. Phys.},
  volume = {98},
  issue = {2},
  pages = {025001},
  numpages = {63},
  year = {2026},
  month = {Apr},
  publisher = {American Physical Society},
  doi = {10.1103/yx73-dk86},
  url = {https://link.aps.org/doi/10.1103/yx73-dk86}
}

@article{claeys2020maximumvelocity,
  title = {Maximum velocity quantum circuits},
  author = {Claeys, Pieter W. and Lamacraft, Austen},
  journal = {Phys. Rev. Res.},
  volume = {2},
  issue = {3},
  pages = {033032},
  numpages = {20},
  year = {2020},
  month = {Jul},
  publisher = {American Physical Society},
  doi = {10.1103/PhysRevResearch.2.033032},
  url = {https://link.aps.org/doi/10.1103/PhysRevResearch.2.033032}
}

@article{preskill2018quantum,
  doi = {10.22331/q-2018-08-06-79},
  url = {https://doi.org/10.22331/q-2018-08-06-79},
  title = {Quantum {C}omputing in the {NISQ} era and beyond},
  author = {Preskill, John},
  journal = {{Quantum}},
  issn = {2521-327X},
  publisher = {{Verein zur F{\"{o}}rderung des Open Access Publizierens in den Quantenwissenschaften}},
  volume = {2},
  pages = {79},
  month = aug,
  year = {2018}
}

@article{fauseweh2024quantum,
  title={Quantum many-body simulations on digital quantum computers: State-of-the-art and future challenges},
  author={Fauseweh, Benedikt},
  journal={Nature Communications},
  volume={15},
  number={1},
  pages={2123},
  year={2024},
  publisher={Nature Publishing Group UK London},
doi={https://doi.org/10.1038/s41467-024-46402-9}
}

@article{ippoliti2021many,
  title = {Many-Body Physics in the {NISQ} Era: Quantum Programming a Discrete Time Crystal},
  author = {Ippoliti, Matteo and Kechedzhi, Kostyantyn and Moessner, Roderich and Sondhi, S.L. and Khemani, Vedika},
  journal = {PRX Quantum},
  volume = {2},
  issue = {3},
  pages = {030346},
  numpages = {24},
  year = {2021},
  month = {Sep},
  publisher = {American Physical Society},
  doi = {10.1103/PRXQuantum.2.030346},
  url = {https://link.aps.org/doi/10.1103/PRXQuantum.2.030346}
}

@article{smith2019simulating,
  title={Simulating quantum many-body dynamics on a current digital quantum computer},
  author={Smith, Adam and Kim, MS and Pollmann, Frank and Knolle, Johannes},
  journal={npj Quantum Information},
  volume={5},
  number={1},
  pages={106},
  year={2019},
  publisher={Nature Publishing Group UK London},
doi={https://doi.org/10.1038/s41534-019-0217-0}

}

@article{hoke2023measurement,
   title={Measurement-induced entanglement and teleportation on a noisy quantum processor},
   volume={622},
   ISSN={1476-4687},
   url={http://dx.doi.org/10.1038/s41586-023-06505-7},
   DOI={10.1038/s41586-023-06505-7},
   number={7983},
   journal={Nature},
   publisher={Springer Science and Business Media LLC},
   author={Hoke, J. C. and Ippoliti, M. and Rosenberg, E. and Abanin, D. and Acharya, R. and Andersen, T. I. and Ansmann, M. and Arute, F. and Arya, K. and others},
   year={2023},
   month=oct, pages={481–486} }

@article{fisher2023random,
   author = "Fisher, Matthew P. A. and Khemani, Vedika and Nahum, Adam and Vijay, Sagar",
   title = "Random Quantum Circuits", 
   journal= "Annual Review of Condensed Matter Physics",
   year = "2023",
   volume = "14",
   number = "Volume 14, 2023",
   pages = "335-379",
   doi = "https://doi.org/10.1146/annurev-conmatphys-031720-030658",
   url = "https://www.annualreviews.org/content/journals/10.1146/annurev-conmatphys-031720-030658",
   publisher = "Annual Reviews",
   issn = "1947-5462",
   type = "Journal Article",
  }

@article{boixo2018characterising,
   title={Characterizing quantum supremacy in near-term devices},
   volume={14},
   ISSN={1745-2481},
   url={http://dx.doi.org/10.1038/s41567-018-0124-x},
   DOI={10.1038/s41567-018-0124-x},
   number={6},
   journal={Nature Physics},
   publisher={Springer Science and Business Media LLC},
   author={Boixo, Sergio and Isakov, Sergei V. and Smelyanskiy, Vadim N. and Babbush, Ryan and Ding, Nan and Jiang, Zhang and Bremner, Michael J. and Martinis, John M. and Neven, Hartmut},
   year={2018},
   month=apr, pages={595–600} }

@article{arute2019quantum,
   title={Quantum supremacy using a programmable superconducting processor},
   volume={574},
   ISSN={1476-4687},
   url={http://dx.doi.org/10.1038/s41586-019-1666-5},
   DOI={10.1038/s41586-019-1666-5},
   number={7779},
   journal={Nature},
   publisher={Springer Science and Business Media LLC},
   author={Arute, Frank and Arya, Kunal and Babbush  {\it et al.}, Ryan },
   year={2019},
   month=oct, pages={505–510} }

@article{chan2018solution,
  title = {Solution of a Minimal Model for Many-Body Quantum Chaos},
  author = {Chan, Amos and De Luca, Andrea and Chalker, J. T.},
  journal = {Phys. Rev. X},
  volume = {8},
  issue = {4},
  pages = {041019},
  numpages = {17},
  year = {2018},
  month = {Nov},
  publisher = {American Physical Society},
  doi = {10.1103/PhysRevX.8.041019},
  url = {https://link.aps.org/doi/10.1103/PhysRevX.8.041019}
}

@article{khemani2018operator,
  title = {Operator Spreading and the Emergence of Dissipative Hydrodynamics under Unitary Evolution with Conservation Laws},
  author = {Khemani, Vedika and Vishwanath, Ashvin and Huse, David A.},
  journal = {Phys. Rev. X},
  volume = {8},
  issue = {3},
  pages = {031057},
  numpages = {25},
  year = {2018},
  month = {Sep},
  publisher = {American Physical Society},
  doi = {10.1103/PhysRevX.8.031057},
  url = {https://link.aps.org/doi/10.1103/PhysRevX.8.031057}
}

@article{nahum2017quantum,
  title = {Quantum Entanglement Growth under Random Unitary Dynamics},
  author = {Nahum, Adam and Ruhman, Jonathan and Vijay, Sagar and Haah, Jeongwan},
  journal = {Phys. Rev. X},
  volume = {7},
  issue = {3},
  pages = {031016},
  numpages = {30},
  year = {2017},
  month = {Jul},
  publisher = {American Physical Society},
  doi = {10.1103/PhysRevX.7.031016},
  url = {https://link.aps.org/doi/10.1103/PhysRevX.7.031016}
}

@article{chan2018spectral,
  title = {Spectral Statistics in Spatially Extended Chaotic Quantum Many-Body Systems},
  author = {Chan, A. and De Luca, A. and Chalker, J. T.},
  journal = {Phys. Rev. Lett.},
  volume = {121},
  issue = {6},
  pages = {060601},
  numpages = {5},
  year = {2018},
  month = {Aug},
  publisher = {American Physical Society},
  doi = {10.1103/PhysRevLett.121.060601},
  url = {https://link.aps.org/doi/10.1103/PhysRevLett.121.060601}
}

@article{keyserlingk2018operator,
  title = {Operator Hydrodynamics, OTOCs, and Entanglement Growth in Systems without Conservation Laws},
  author = {von Keyserlingk, C. W. and Rakovszky, Tibor and Pollmann, Frank and Sondhi, S. L.},
  journal = {Phys. Rev. X},
  volume = {8},
  issue = {2},
  pages = {021013},
  numpages = {19},
  year = {2018},
  month = {Apr},
  publisher = {American Physical Society},
  doi = {10.1103/PhysRevX.8.021013},
  url = {https://link.aps.org/doi/10.1103/PhysRevX.8.021013}
}

@article{rakovszky2018diffusive,
  title = {Diffusive Hydrodynamics of Out-of-Time-Ordered Correlators with Charge Conservation},
  author = {Rakovszky, Tibor and Pollmann, Frank and von Keyserlingk, C. W.},
  journal = {Phys. Rev. X},
  volume = {8},
  issue = {3},
  pages = {031058},
  numpages = {28},
  year = {2018},
  month = {Sep},
  publisher = {American Physical Society},
  doi = {10.1103/PhysRevX.8.031058},
  url = {https://link.aps.org/doi/10.1103/PhysRevX.8.031058}
}

@article{roberts2015diagnosing,
  title = {Diagnosing Chaos Using Four-Point Functions in Two-Dimensional Conformal Field Theory},
  author = {Roberts, Daniel A. and Stanford, Douglas},
  journal = {Phys. Rev. Lett.},
  volume = {115},
  issue = {13},
  pages = {131603},
  numpages = {6},
  year = {2015},
  month = {Sep},
  publisher = {American Physical Society},
  doi = {10.1103/PhysRevLett.115.131603},
  url = {https://link.aps.org/doi/10.1103/PhysRevLett.115.131603}
}

@article{maldacena2016bound,
   title={A bound on chaos},
   volume={2016},
   ISSN={1029-8479},
   url={http://dx.doi.org/10.1007/JHEP08(2016)106},
   pages={106},
   journal={Journal of High Energy Physics},
   publisher={Springer Science and Business Media LLC},
   author={Maldacena, Juan and Shenker, Stephen H. and Stanford, Douglas},
   year={2016},
   month={Aug} }

@article{bohrdt2017scrambling,
    doi = {10.1088/1367-2630/aa719b},
    url = {https://doi.org/10.1088/1367-2630/aa719b},
    year = 2017,
    month = {jun},
    publisher = {{IOP} Publishing},
    volume = {19},
    number = {6},
    pages = {063001},
    author = {A Bohrdt and C B Mendl and M Endres and M Knap},
    title = {Scrambling and thermalization in a diffusive quantum many-body system},
    journal = {New J. Phys.}
}

@article{luitz2017information,
  title = {Information propagation in isolated quantum systems},
  author = {Luitz, David J. and Bar Lev, Yevgeny},
  journal = {Phys. Rev. B},
  volume = {96},
  issue = {2},
  pages = {020406},
  numpages = {5},
  year = {2017},
  month = {Jul},
  publisher = {American Physical Society},
  doi = {10.1103/PhysRevB.96.020406},
  url = {https://link.aps.org/doi/10.1103/PhysRevB.96.020406}
}

@article{bilitewski2021classical,
  title = {Classical many-body chaos with and without quasiparticles},
  author = {Bilitewski, Thomas and Bhattacharjee, Subhro and Moessner, Roderich},
  journal = {Phys. Rev. B},
  volume = {103},
  issue = {17},
  pages = {174302},
  numpages = {25},
  year = {2021},
  month = {May},
  publisher = {American Physical Society},
  doi = {10.1103/PhysRevB.103.174302},
  url = {https://link.aps.org/doi/10.1103/PhysRevB.103.174302}
}

\clearpage

\section{End Matter}
\subsection{Rules for the Markov process}
To lay out the rules for the Markov process in detail, it is useful to define the notation
\eq{
    \Gamma_{\cal S}\equiv\{\eta_{x,\cal S}^\mu\}\equiv  \cdots\underbrace{\bullet\circ}_{x-1}~\underbrace{\bullet\bullet}_x~\underbrace{\circ\circ}_{x+1}\cdots\,,
    \label{eq:string-pic}
}
where the two circles at a site denote the two Majorana slots, with filled and empty denoting the corresponding Majorana to be present and absent, respectively, in the configuration. In the model in Eq.~\eqref{eq:brickwork}, each gate acting on neighbouring sites, $x$ and $x+1$, is either a matchgate, denoted by $\cbox{\begin{tikzpicture}[scale=0.5]
\draw (0,-0.4) -- (0, 0.4);
\draw (1,-0.4) -- (1, 0.4);
\filldraw[draw=black,fill=blue!20] (0-0.1,0-0.25) rectangle (0+1.1,0+0.25);
\end{tikzpicture}}$, or a $\swap$ gate sandwiched between two matchgates, denoted by  $\cbox{\begin{tikzpicture}[scale=0.5]
\draw (0,-0.4) -- (0, 0.4);
\draw (1,-0.4) -- (1, 0.4);
\filldraw[draw=black,fill=red!20] (0-0.1,0-0.25) rectangle (0+1.1,0+0.25);
\fill[fill=blue!40] (0-0.1,0-0.25) rectangle (0+1.1,-0.125);
\fill[fill=blue!40] (0-0.1,0.125) rectangle (0+1.1,0.25);
\end{tikzpicture}}$.
Averaging over the gates, as mentioned in the main text, the transition probability from a string $\Gamma_{\cal S}$ to another, $\Gamma_{{\cal S}'}$, is given by
\eq{
p(\Gamma_{\cal S} \to \Gamma_{{\cal S}'}) = \langle|{\rm tr}[\Gamma_{{\cal S}'}U_{x,x+1}\Gamma_{\cal S} U_{x,x+1}^\dagger|^2\rangle_{M}\,.
}
We  denote this pictorially as $\cbox{\begin{tikzpicture}[scale=0.5]
% \draw (1+0.2,-0.8+0.1) -- (1+0.2, 0.8+0.1);
% \draw (0.2,-0.8+0.1) -- (0.2, 0.8+0.1);
% \filldraw[draw=black,fill=gray!50] (0-0.1+0.2,0-0.25+0.1) rectangle (0+1.1+0.2,0+0.25+0.1);
\draw (0,-0.8) -- (0, 0.4);
\draw (1,-0.8) -- (1, 0.4);
\filldraw[draw=black,fill=white] (0-0.1,0-0.25) rectangle (0+1.1,0+0.25);
% \filldraw[draw=black,fill=Blue, rounded corners=.09cm] (0-0.1,0.4) rectangle (0+1.1,0.65);
\filldraw[draw=black,fill=YellowOrange, rounded corners=.07cm] (0-0.1,-0.4) rectangle (0+1.1,-0.65);
\node at (1.7,-0.6) {{\color{YellowOrange}{\small $\Gamma_{\cal S}$}}};
\node at (2,0) {$\rightarrow$};
\draw (3,-0.8) -- (3, 0.);
\draw (4,-0.8) -- (4, 0.);
\filldraw[draw=black,fill=OliveGreen, rounded corners=.07cm] (3-0.1,-0.4) rectangle (3+1.1,-0.65);
\node at (4.7,-0.6) {{\color{OliveGreen}{\small $\Gamma_{{\cal S}'}$}}};
% \draw (0,0.8) to[out=90,in=90] (0.2,0.9);
% \draw (1,0.8) to[out=90,in=90] (1.2,0.9);
% \draw (0,-0.8) to[out=-90,in=-90] (0.2,-0.7);
% \draw (1,-0.8) to[out=-90,in=-90] (1.2,-0.7);
\end{tikzpicture}}$, where the white gate could be either the matchgate or the $\swap$ gate as described above. 
Using this notation and the one described in Eq.~\eqref{eq:string-pic}, the rules of the Markov process can be summarised as 
\begin{widetext}
\eq{
    \begin{split}
    &\cbox{\begin{tikzpicture}[scale=0.5]
\draw (0,-0.4) -- (0, 0.4);
\draw (1,-0.4) -- (1, 0.4);
\filldraw[draw=black,fill=blue!20] (0-0.1,0-0.25) rectangle (0+1.1,0+0.25);
\node at (0,-0.5) {$\bullet \circ$};
\node at (1,-0.5) {$\circ \circ$};
\end{tikzpicture}}\to \begin{cases}
                                        \bullet\!\circ~\circ\circ;~&p=1/4\\
                                        \circ\!\bullet~\circ\circ;~&p=1/4\\
                                        \circ\!\circ~\bullet\circ;~&p=1/4\\
                                        \circ\!\circ~\circ\bullet;~&p=1/4
                                     \end{cases}\,;~
    \cbox{\begin{tikzpicture}[scale=0.5]
\draw (0,-0.4) -- (0, 0.4);
\draw (1,-0.4) -- (1, 0.4);
\filldraw[draw=black,fill=blue!20] (0-0.1,0-0.25) rectangle (0+1.1,0+0.25);
\node at (0,-0.5) {$\bullet \bullet$};
\node at (1,-0.5) {$\bullet \circ$};
\end{tikzpicture}}\to \begin{cases}
                                        \bullet\!\bullet~\bullet\circ;~&p=1/4\\
                                        \bullet\!\bullet~\circ\bullet;~&p=1/4\\
                                        \bullet\!\circ~\bullet\bullet;~&p=1/4\\
                                        \circ\!\bullet~\bullet\bullet;~&p=1/4
                                     \end{cases}\,;~
    \cbox{\begin{tikzpicture}[scale=0.5]
\draw (0,-0.4) -- (0, 0.4);
\draw (1,-0.4) -- (1, 0.4);
\filldraw[draw=black,fill=blue!20] (0-0.1,0-0.25) rectangle (0+1.1,0+0.25);
\node at (0,-0.5) {$\bullet \circ$};
\node at (1,-0.5) {$\bullet \circ$};
\end{tikzpicture}},\cbox{\begin{tikzpicture}[scale=0.5]
\draw (0,-0.4) -- (0, 0.4);
\draw (1,-0.4) -- (1, 0.4);
\filldraw[draw=black,fill=red!20] (0-0.1,0-0.25) rectangle (0+1.1,0+0.25);
\fill[fill=blue!40] (0-0.1,0-0.25) rectangle (0+1.1,-0.125);
\fill[fill=blue!40] (0-0.1,0.125) rectangle (0+1.1,0.25);
\node at (0,-0.5) {$\bullet \circ$};
\node at (1,-0.5) {$\bullet \circ$};
\end{tikzpicture}}\to \begin{cases}
                                        \bullet\!\bullet~\circ\circ;~&p=1/6\\
                                        \circ\!\bullet~\bullet\circ;~&p=1/6\\
                                        \circ\!\bullet~\circ\bullet;~&p=1/6\\
                                        \bullet\!\circ~\circ\bullet;~&p=1/6\\
                                        \bullet\!\circ~\bullet\circ;~&p=1/6\\
                                        \circ\!\circ~\bullet\bullet;~&p=1/6
                                     \end{cases}\,;\\
    &\cbox{\begin{tikzpicture}[scale=0.5]
\draw (0,-0.4) -- (0, 0.4);
\draw (1,-0.4) -- (1, 0.4);
\filldraw[draw=black,fill=red!20] (0-0.1,0-0.25) rectangle (0+1.1,0+0.25);
\fill[fill=blue!40] (0-0.1,0-0.25) rectangle (0+1.1,-0.125);
\fill[fill=blue!40] (0-0.1,0.125) rectangle (0+1.1,0.25);
\node at (0,-0.5) {$\bullet \circ$};
\node at (1,-0.5) {$\circ \circ$};
\end{tikzpicture}}\to \begin{cases}
                                        \bullet\!\bullet~\bullet\circ;~&p=1/4\\
                                        \bullet\!\bullet~\circ\bullet;~&p=1/4\\
                                        \bullet\!\circ~\bullet\bullet;~&p=1/4\\
                                        \circ\!\bullet~\bullet\bullet;~&p=1/4
                                     \end{cases}\,;~
    \cbox{\begin{tikzpicture}[scale=0.5]
\draw (0,-0.4) -- (0, 0.4);
\draw (1,-0.4) -- (1, 0.4);
\filldraw[draw=black,fill=red!20] (0-0.1,0-0.25) rectangle (0+1.1,0+0.25);
\fill[fill=blue!40] (0-0.1,0-0.25) rectangle (0+1.1,-0.125);
\fill[fill=blue!40] (0-0.1,0.125) rectangle (0+1.1,0.25);
\node at (0,-0.5) {$\bullet \bullet$};
\node at (1,-0.5) {$\bullet \circ$};
\end{tikzpicture}}\to \begin{cases}
                                        \bullet\!\circ~\circ\circ;~&p=1/4\\
                                        \circ\!\bullet~\circ\circ;~&p=1/4\\
                                        \circ\!\circ~\bullet\circ;~&p=1/4\\
                                        \circ\!\circ~\circ\bullet;~&p=1/4
                                     \end{cases}\,;~
    \cbox{\begin{tikzpicture}[scale=0.5]
\draw (0,-0.4) -- (0, 0.4);
\draw (1,-0.4) -- (1, 0.4);
\filldraw[draw=black,fill=blue!20] (0-0.1,0-0.25) rectangle (0+1.1,0+0.25);
\node at (0,-0.5) {$\bullet \bullet$};
\node at (1,-0.5) {$\bullet \bullet$};
\end{tikzpicture}},\cbox{\begin{tikzpicture}[scale=0.5]
\draw (0,-0.4) -- (0, 0.4);
\draw (1,-0.4) -- (1, 0.4);
\filldraw[draw=black,fill=red!20] (0-0.1,0-0.25) rectangle (0+1.1,0+0.25);
\fill[fill=blue!40] (0-0.1,0-0.25) rectangle (0+1.1,-0.125);
\fill[fill=blue!40] (0-0.1,0.125) rectangle (0+1.1,0.25);
\node at (0,-0.5) {$\bullet \bullet$};
\node at (1,-0.5) {$\bullet \bullet$};
\end{tikzpicture}}\to \bullet\!\bullet~\bullet\bullet;~p=1\,;\\
    \end{split}\,.
}
\label{eq:markov-rules}
\end{widetext}
The rules above completely specify the Markov process, which can be simulated readily for arbitrarily large systems, in turn yielding the averaged OTOCs.

\subsection{Additional numerical results for the crossover regime}
\begin{figure}[!b]
\includegraphics[width=\linewidth]{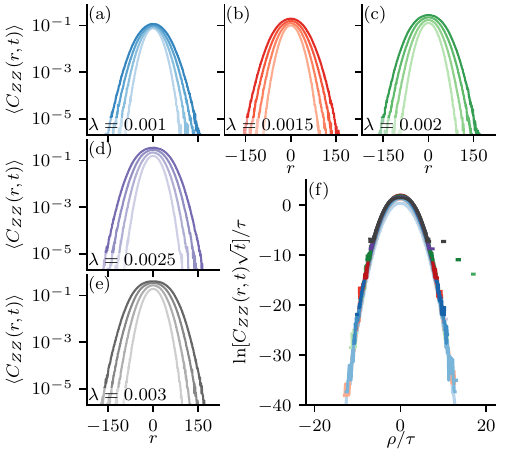}
\caption{(a)-(e) The spatial profile of the OTOC as a function of $r$ for different times $t=200,300,400,500$ (lighter to darker colours) for different values of $\lambda$ (different panels). (f) All the data shown on panels (a)-(e) collapsed onto a common curve following the crossover scaling form of the OTOC in Eq.~\eqref{eq:OTOC-crossover-scaling}. }
\label{fig:crossover-space}
\end{figure}

In Fig.~\ref{fig:crossover}, we provided evidence for the scaling form in Eq.~\eqref{eq:OTOC-crossover-scaling} by considering two representative spatial cuts and plotting the OTOC as a function of rescaled $t$ for several values of $\lambda$.
Here we take the complementary approach and show the entire spatial profile of the OTOC as a function of the rescaled spatial coordinate for several values of $\lambda$ and at a few representative temporal cuts.
The results, shown in Fig.~\ref{fig:crossover-space}, again confirm that 
\eq{
    \ln[\braket{C_{ZZ}(r,t)}\sqrt{t}] = \tau g(\rho/\tau)\,,
}
where $\tau = t/t_\ast(\lambda)$ and $\rho = r/r_\ast(\lambda)$, with ${t_\ast(\lambda)\sim \lambda^{-1}}$, and ${r_\ast(\lambda)\sim \lambda^{-1/2}}$. 
This is exactly the scaling form in Eq.~\eqref{eq:OTOC-crossover-scaling}.

\subsection{Additional details of the noisy F-KPP equation}
\subsubsection{Derivation of the non-linearities and noise}

To derive the non-linear terms induced by the $\swap$ gates in the F-KPP equation \eqref{eq:FKPP} at the mean-field level, we assume local `thermalisation' of the operator strings.
Within this assumption, the probability of occupying a Majorana slot at $(r,t)$ is ${\cal C}(r,t)$ and the probability of it being empty is $1-{\cal C}(r,t)$.
Under the branching process $\bullet\!\circ~\circ\circ\to\bullet\!\bullet~\bullet\circ$, the OTOC effectively increases by 2.
Similarly, under the annihilation process,
$\bullet\!\bullet~\bullet\circ\to\bullet\!\circ~\circ\circ$, the OTOC decreases by 2.
The probability of having a single Majorana slot occupied out of the four slots on two sites is $p_1 = 4 {\cal C}(r,t)[1-{\cal C}(r,t)]^3$ and the probability of having three of them occupied is $p_3 = 4{\cal C}^3(r,t)[1-{\cal C}(r,t)]$.
The net change in the average OTOC under the $\swap$ gate is then
\eq{
\braket{\Delta {\cal C}(r,t)}&= \frac{1}{4}[2\times p_1 - 2\times p_3]\nonumber\\
&= 2 {\cal C}(r,t)[1-{\cal C}(r,t)][1-2{\cal C}(r,t)]\,,\label{eq:Delta-swap}
}
where the factor of $1/4$ just encodes the uniform distribution of the OTOC across the four Majorana slots by the matchgate.
The expression in Eq.~\eqref{eq:Delta-swap} is precisely the non-linear term, $H[\cal C]$, in Eq.~\eqref{eq:FKPP}. 

The noise term can also be derived in a similar spirit. 
While Eq.~\eqref{eq:Delta-swap} encodes the average change in the OTOC due to the $\swap$ gate, including the fluctuations amounts to writing
\eq{
\Delta {\cal C}(r,t) = \braket{\Delta {\cal C}(r,t)}
+ \vartheta(r,t)\eta(r,t)\,,
}
where 
$\eta(r,t)$ denotes white noise and $\vartheta(r,t)$ can be obtained as 
\eq{
\vartheta(r,t) &= \sqrt{\frac{1}{4^2}[(2)^2 \times p_1 + (-2)^2 \times p_3]} \nonumber\\&= \sqrt{{\cal C}[1-{\cal C}][1-2{\cal C}+2{\cal C}^2]}\,,
\label{eq:noise-demo}
}
which is exactly the noise term, $\sqrt{K[\cal C]}$, in Eq.~\eqref{eq:FKPP}.

\subsubsection{Absence of $\lambda$-dependence on diffusive broadening}

To show that the diffusion constant associated to broadening of the OTOC front has a negligible dependence on $\lambda$, it is useful to write Eq.~\eqref{eq:FKPP} in transformed coordinates, $\rho$ and $\tau$ as
\eq{
    \frac{1}{t_\ast}\partial_\tau{\cal C} = \frac{D}{r_\ast^2}\partial_\rho^2{\cal C} +\lambda H[{\cal C}] + \frac{1}{\sqrt{r_\ast t_\ast}}\sqrt{\lambda K[{\cal C}]}\tilde{\eta}(\rho,\tau)\,.
}
Using the fact that $r_\ast\sim \lambda^{-1/2}$ and $t_\ast\sim \lambda^{-1}$, the above equation can be recast as 
\eq{
    \partial_\tau{\cal C} = a_0D\partial_\rho^2{\cal C} +a_1 H[{\cal C}] + \epsilon(\lambda)\sqrt{ K[{\cal C}]}\tilde{\eta}(\rho,\tau)\,,
    \label{eq:non-dim}
}
where $a_0,a_1$ are constants and $\epsilon(\lambda)\sim \lambda^{1/4}$.
Note that in the above equation, none of the terms except the noise term depends on $\lambda$.
Therefore the broadening of the front in the rescaled coordinates is given by
\eq{
\braket{\delta \rho_{\rm front}^2}\sim D_{\rm eff}(\epsilon)\tau\,.
}
Transforming back to the regular coordinates we have
\eq{
\frac{1}{r_\ast^2}\braket{\delta r^2_{\rm front}}\sim D_{\rm eff}(\epsilon)\frac{t}{t_\ast}\Rightarrow \braket{\delta r^2_{\rm front}}\sim D_{\rm eff}(\epsilon)t\,,
}
where we again used the scalings of $r_\ast$ and $t_\ast$ with $\lambda$.
This shows that the diffusion constant of the broadening of the front is indeed $\sim D_{\rm eff}(\epsilon)$. 
It was shown quite generally in Refs.~\cite{brunet2006phenomenological} that for equations of the form \eqref{eq:non-dim},
$D_{\rm eff}(\epsilon) \sim D/|\ln \epsilon|^3$. 
Since $\epsilon\sim \lambda^{1/4}$, the diffusion constant only has a very weak logarithmic dependence on $\lambda$.
From a microscopic point of view, this negligible dependence on $\lambda$ of the front broadening can be understood as a competition (and cancellation) between two effects.
On increasing $\lambda$, the intrinsic non-linearity of the dynamics increases which amplifies the birth rate of the OTOC at the leading edge and tries to push the front further out.
At the same time, on increasing $\lambda$, the front becomes steeper due to larger rates of birth processes in the interior which results in a suppression of the width of the front restricting how far a random fluctuation can push out the front.

\end{document}